\documentclass[12pt]{article}

\renewcommand{\baselinestretch}{1.5}
\long\def\symbolfootnote[#1]#2{\begingroup
\def\thefootnote{\fnsymbol{footnote}}\footnote[#1]{#2}\endgroup} 

\usepackage[margin=.9in]{geometry}
\usepackage{enumerate}
\usepackage{amsmath,graphicx,array,hyperref} 
\usepackage{multirow}

 \newcommand{\iid}{\stackrel{\mathrm{iid}}{\sim}}

\usepackage{tikz}
\usetikzlibrary{arrows}
\usepackage{amstext,amsthm, amssymb,amsfonts}
\usepackage{natbib,bbm,dsfont}

\begin{document}

\title{Testing and Modeling Dependencies Between \\a Network and Nodal Attributes}
\author{Bailey K. Fosdick$^1$ and Peter D. Hoff$^{1,2}$ \\ 
  Departments of Statistics$^1$ and Biostatistics$^2$, University of Washington}
\date{\today}
\maketitle

\symbolfootnote[0]{This work was partially supported by NICHD grant R01 HD-067509.  The authors thank Alex Volfovsky for his helpful comments and discussion.}

\renewcommand{\baselinestretch}{1.25}
\begin{abstract}
\medskip  
Network analysis is often focused on characterizing the dependencies between network relations and node-level attributes.  Potential relationships are typically explored by modeling the network as a function of the nodal attributes or by modeling the attributes as a function of the network.  These methods require specification of the exact nature of the association between the network and attributes, reduce the network data to a small number of summary statistics, and are unable provide predictions simultaneously for missing attribute and network information.  Existing methods that model the attributes and network jointly also assume the data are fully observed.  In this article we introduce a unified approach to analysis that addresses these shortcomings.  We use a latent variable model to obtain a low dimensional representation of the network in terms of node-specific network factors and use a test of dependence between the network factors and attributes as a surrogate for a test of dependence between the network and attributes.  We propose a formal testing procedure to determine if dependencies exists between the network factors and attributes.  We also introduce a joint model for the network and attributes, for use if the test rejects, that can capture a variety of dependence patterns and be used to make inference and predictions for missing observations.   \\

\noindent {\textit{Keywords:} hypothesis test; joint model; latent variable model; prediction; relational data} 
\end{abstract}

\renewcommand{\baselinestretch}{1.5}
\section{Introduction}
A common goal in the analysis of network data is to characterize the dependence between network relations and a set of node-specific attributes.  For example in recent years many studies in the social sciences have examined the relationship between individuals' friendship networks and their health measures, such as happiness (\cite{FowlerChristakis2008}), smoking and drinking behavior (\cite{Kiuru2010}), and obesity (\cite{ChristakisFowlder2007}, 
\cite{delaHaye2010}).  Similarly in the biological sciences, scientists are interested in the relationship between how proteins interact and their biological importance (see \cite{Butland} for example).  In each of these applications, the data consists of two parts: the network relations $ \{y_{i,j} : i,j \in \{1,...,n\} \}$ representing a measure of the directed relationship between each pair of nodes $i$ and $j$, and $p$-variate nodal attributes $ \{\boldsymbol{x}_{i} : i \in \{1,...,n\} \}$.  In the case of a social network, the nodes, network relations, and attributes 
  often represent people, their friendships, and their demographic and behavioral characteristics, respectively. 

Traditional approaches to describing the dependence between a network and attributes rely on statistical methods that model either the network conditional on the attributes or the attributes conditional on the network.  In the social sciences, this first perspective parallels the theory of ``social selection'', whereby individuals' attributes influence the formation of their social relations, and the second perspective is motivated by ``social influence'' theory whereby individuals' relations affect their attributes.   

Methods that model the network as a function of the attributes commonly specify a regression framework for the dependence: the probability of the relation $y_{i,j}$ is a function of $\boldsymbol{\beta}^{T}\boldsymbol{x}_{i,j}$ where $\boldsymbol{x}_{i,j} = f(\boldsymbol{x}_{i},\boldsymbol{x}_{j})$, $\boldsymbol{x}_{i}$ and $\boldsymbol{x}_{j}$ are the attributes for nodes $i$ and $j$, and $\boldsymbol{\beta}$ is an unknown parameter vector.   The covariate vector $\boldsymbol{x}_{i,j}$ typically includes terms for each attribute of the sender node $i$ and receiver node $j$, as well as interaction terms measuring the similarity between the sender and receiver attributes.  These interaction terms are frequently defined as the absolute difference between the attributes, an indicator of whether an attribute is the same for both the sender and receiver node (in the case of discrete attributes), or the product of the nodes' attributes.  Examples of network models that can accommodate such a regression term are exponentially parameterized random graph models (ERGM) (\cite{FrankStrauss}, \cite{WassermanPattison1996}, \cite{Snijders2006}, \cite{HunterHandcock2006}) 
and latent variables models (\cite{HoffRafteryHandcock2002}, \cite{Hoff2005}).  This latter class of models regresses a function of the network on both attribute terms and node-specific latent variables;  \cite{Austin2013} proposed a slight modification to this class where the network is expressed as a function of nodal latent variables and the latent variables are regressed on the attributes.  

Methods for assessing the impact of the network on nodal attributes often regress each node's attributes on the attributes of other nodes in the network according to the network relations.  For example, \cite{ChristakisFowlder2007} use a logistic regression model to estimate the degree to which an individual's obesity status can be explained by the obesity status of individuals in their social network (children, neighbors, spouse, etc.).   Other similar models include the auto-regressive network effects models of \cite{ErbringYoung} and \cite{MarsdenFriedkin} and the p$^{*}$ social influence models of \cite{RobinsPattisonElliott2001influence}.  All of these models are univariate, focusing a single attribute of interest that is possibly subject to social influence.

While modeling the network and attributes as functions of one another is able to provide some insight into their dependence structure, there are two primary drawbacks to utilizing these methods for analysis.  First, neither modeling framework allows for simultaneous inference about the dependencies between and among the network relations and attributes.  For example, when analyzing data on an adolescent friendship network and individuals' health behaviors there may be interest in whether smoking habits and obesity status are conditionally independent given the network.  Addressing this question of dependence between attributes conditional on the network is impossible using either of the conditional modeling frameworks.
A second limitation of these methods is that they are unable to accommodate, and provide predictions for, datasets that have both missing network and attribute information.  In the conditional modeling frameworks either the network or attributes are assumed to be fully observed.

\cite{FellowsHandcock} proposed a new class of models called exponential-family random network models which is a combination of an ERGM and a Gibbs random field.  This joint network and attribute model addresses the first limitation of the conditional models and could potentially (with modification) address the second limitation of imputing missing data.  However these models, like ERGMs, are difficult to estimate and can suffer from model degeneracy problems, where networks simulated from the fitted model are unlike that which was observed (\cite{Handcock2003}, \cite{Schweinberger2011}).  \cite{Kim2011} and \cite{Kim2012} proposed a simple joint attribute and network model for which mathematical analysis on network connectivity and degree distributions is tractable.  However, this model class only accommodates categorical attributes and assumes no missing attribute or network information.  Both of these existing joint modeling frameworks lack traditional procedures for testing whether the joint model is appropriate and dependencies even exist between the network and attributes.  

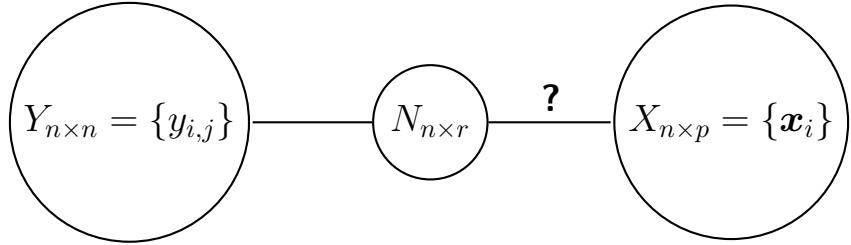
\begin{figure}
\begin{center}
\begin{tikzpicture}[-,>=stealth',shorten >=1pt,auto,node distance=4cm,  
  thick,main node/.style={circle,draw,font=\sffamily\large\bfseries}]
  \node[main node] (1) {$N_{n \times r}$};
  \node[main node] (2) [left of=1] {$Y_{n \times n} = \{y_{i,j}\}$};
  \node[main node] (3) [right of=1] {$X_{n \times p} = \{\boldsymbol{x}_{i}\}$};

  \path[every node/.style={font=\sffamily\large\bfseries}]
    (1)  edge node[left] {}(2)
       edge node  {?} (3);
\end{tikzpicture}
\end{center}
\label{conceptpic}
\caption{The primary patterns in the network $Y$ are represented by $r$ node-specific factors $N$.  To determine if dependencies exist between the network $Y$ and the $p$ nodal attributes $X$, we propose testing for a the relationship between the network factors and attributes. }
\end{figure}

In this article, we propose a unified approach to the analysis of network and attribute data.  This approach allows for testing for dependencies between the network and attributes, and in the event the test concludes such dependencies exists, jointly modeling the network and attributes to make inference and obtain predictions for missing values.  Our proposed methodology can be summarized as follows.  
Investigating the dependence between network data $Y$ and attribute data $X$ is difficult since network data is often high dimensional, containing relational information on each pair of nodes, and there lacks a one-to-one correspondence  between nodes' network relations and their attributes.  For these reasons, in Section 2 we propose representing the $(n \times n)$ matrix of network relations $Y$ with a low dimensional structure defined by an $(n \times r)$ matrix $N$ of node-specific network factors ($ r \ll n$).  These network factors $N$ are not observed directly and hence are estimated from the observed network $Y$ using a network model.   

In Section 3 we propose evaluating whether dependencies exist between the network $Y$ and attributes $X$ by formally testing for correlation between the estimated network factors $N$ and the attributes $X$.  A conceptual representation of this testing framework is shown in Figure \ref{conceptpic}.  If the network is independent of the attributes, then any functions of the network, specifically the network factors, are also independent of the network.  Therefore, any test of association between the network factors and attributes will have the correct Type I error rate.  A key advantage of this approach is that the overall relationship between the network and an arbitrary number of attributes can be assessed without needing to construct a complex regression model of the network relations on the attributes or perform variable selection.  In Section 4 we investigate the loss in power for the test between the network factors and attributes as a result of not observing the network factors directly. 
  
 A joint model for the network and attributes is presented in Section 5 for use when the test of independence between the network factors and attributes rejects.  This joint model allows for simultaneous estimation and inference on the dependence between and within the network and attributes, as well as provides methodology for handling and predicting missing network and attribute data.  We show that the joint model conditional on the attributes can be viewed as a reduced rank regression of the network relations on attribute interactions.  This further motivates the model as a mechanism for parsimoniously characterizing attribute and network dependence.  In Section \ref{sec:app} our proposed methodology in used to analyze data from the National Longitudinal Study of Adolescent Health.  In a cross validation experiment, we demonstrate that predictions of missing attribute data can be improved by basing imputations on both observed network and attribute information instead of attribute data alone.  We conclude with a discussion in Section \ref{sec:dis}.

\section{Calculation of node-specific network factors}\label{sec:redrep}
The latent space network models in \cite{HoffRafteryHandcock2002} and latent variable models in \cite{Hoff2005} and \cite{Hoff2009} provide parsimonious representations of the patterns in a network using node-specific latent factors.  These models 
 have been shown to capture a variety of network dependence patterns such as homophily, transitivity, reciprocity, and heterogeneity in node sociability and popularity.  We consider an extension of the model presented in \cite{Hoff2009} that contains additive and multiplicative latent effects, as well as structure for within dyad correlation.  Ultimately, we use this model to obtain a low-dimensional representation of the network in terms of interpretable node-specific factors.  We describe the model for continuous network data, however at the end of this section we briefly discuss how these methods can be extended to model ordinal or binary relations.  

Let $y_{i,j}$ represent a continuous measure of the directed relation between nodes $i$ and $j$ and consider the following model:
\begin{align}
y_{i,j} = \mu + a_i + b_j + \boldsymbol{u}_{i}^{T}\boldsymbol{v}_{j} + e_{i,j},  \hspace{.5in} a_{i},b_{j} \in \mathbb{R}, \hspace{.5in} \boldsymbol{u}_{i},\boldsymbol{v}_{j} \in \mathbb{R}^{k}. \label{amen}
\end{align}
The overall mean relation is represented by $\mu$ and the random error by $e_{i,j}$. The additive sender effect $a_{i}$ and receiver effect $b_{j}$ are often interpreted as a measure of node $i$'s sociability (i.e. outgoingness) and node $j$'s popularity respectively.  The multiplicative interaction effect $\boldsymbol{u}_{i}^{T}\boldsymbol{v}_{j}$ can capture higher order dependence, such as network transitivity, balance, and clustering (\cite{Hoff2005}).  One interpretation of the these effects comes from the concept of an underlying social space (\cite{McFarland1973}, \cite{Faust1988}), whereby nodes that are close to one another in the underlying space exhibit similar network patterns.  In this context, the node-specific sender factors $\boldsymbol{u}_{i}$ and receiver factors $\boldsymbol{v}_{i}$ can be interpreted as $k$-dimensional representations of the underlying outgoing (sending) and incoming (receiving) behaviors of node $i$.  A similar interpretation was used to motivate the latent position models in \cite{HoffRafteryHandcock2002}.

The random errors are modeled as Gaussian, independent across dyads, and correlated within a dyad: 
\begin{align}
(e_{i,j},e_{j,i})^T& \iid \text{ normal}_{2} \Big(\boldsymbol{0},\sigma^{2}_{e}\Big(\begin{smallmatrix}
1&\rho\\ \rho&1
\end{smallmatrix} \Big) \Big). 
  \label{err}
\end{align}
The additive and multiplicative node-specific factors are also 
modeled as Gaussian and independent across nodes:
\begin{align}
(a_i, b_i,  \boldsymbol{u}_{i}^{T},\boldsymbol{v}_{i}^{T})^T &\iid \text{ normal}_{2+2k}(\boldsymbol{0},\Sigma_{abuv}) \hspace{.5in}  \Sigma_{abuv} =  \left( \begin{array}{cc}
\Sigma_{ab} & \Sigma_{ab,uv}\\
\Sigma_{uv,ab} & \Sigma_{uv} \\
 \end{array} \right). \label{covABUV}
\end{align}
The within dyad correlation $\rho$ is interpreted as a measure of network relation reciprocity and together with the additive effects $a_{i}$ and $b_{j}$ induces the covariance structure from the social relations model (\cite{SRM}, \cite{Wong1982}).

A Bayesian estimation procedure for this network model has been implemented in the `amen' package in the open source computing software program \texttt{R}; however the implementation restricts $\Sigma_{ab,uv} =0$.  Under this restriction the model can capture third-order dependence patterns between relation ``cycles'', such as $\{y_{i,j},y_{j,k},y_{k,i}\}$ or $\{y_{i,j},y_{j,k},y_{i,k}\}$ where the edges create a closed loop (ignoring direction), but not between noncyclic relation triples such as $\{y_{i,j},y_{j,i},y_{k,i}\}$.  By allowing the additive and multiplicative effects to be dependent (i.e. $\Sigma_{ab,uv} \not=0$) as in \eqref{covABUV}, the model is able to capture a larger class of dependencies.  Specifically, it can capture correlation among sets of relation triples where each relation in the set shares at least one node with another relation in the set (i.e. dependence between $\{y_{i,j},y_{j,k},y_{k,l}\}$, but not between $\{y_{i,j},y_{j,k},y_{m,l}\}$).  One justification for allowing such dependence is that latent factors that act additively, affecting node popularity and sociability, plausibly also impact the network in a multiplicative manner.  
A modified version of the `amen' \texttt{R} package that supports Bayesian parameter estimation for the network model presented here is available at the corresponding author's website.\\

\noindent \textbf{Motivation via the singular value decomposition}\\
A key strength of the network model in \eqref{amen} is its ability to capture a variety of common network phenomena, however an alternative motivation for the model stems from its relationship to the singular value decomposition (SVD).  The singular value decomposition is a matrix factorization that is commonly used to obtain an approximation of a matrix $M$ by another matrix $\widehat{M}$ which is of reduced rank and contains the main patterns of the original matrix $M$.  The SVD-based approximation $\widehat{M}$ is the optimal matrix approximation of its rank with respect to squared error loss.  Here we show that the model in \eqref{amen} is similar to an SVD-based approximation of the network $Y$, and hence can be viewed as a low dimensional representation of the network that captures the primary patterns in the relations.

The network model in \eqref{amen} can be written in matrix form as $Y = M + E $, where
 \begin{align}
M&=\mu\boldsymbol{1_{n}}\boldsymbol{1_{n}}^T +  \boldsymbol{a}\boldsymbol{1_{n}}^T + \boldsymbol{1_{n}}\boldsymbol{b}^T + UV^{T}, \label{amenMATRIX}
\end{align}
$\boldsymbol{a}$ and $\boldsymbol{b}$ are $(n \times 1)$ vectors of the additive sender and receiver factors, $U$ and $V$ are $(n \times k)$ matrices of multiplicative factors, and $E$ is an $(n \times n)$ matrix of errors.  

The singular value decomposition of an arbitrary $(n \times n)$ matrix $Y$ is written $Y=ACB^T$, where $A$ and $B$ are orthogonal $(n \times n)$ matrices and $C$ is an $(n \times n)$ diagonal matrix with non-negative decreasing diagonal elements.  The rank-$k$ matrix that best approximates $Y$ based on squared-error loss is given by $\widehat{M} = \widehat{A}\widehat{C}\widehat{B}^T$ where $\widehat{A} = A[,1 \colon k]$, $\widehat{C} = C[1 \colon k,1\colon k]$ and $\widehat{B} =B[,1 \colon k]$.  Absorbing $\widehat{C}$ into $\widehat{A}$ and/or $\widehat{B}$, the best rank-$k$ approximation is written $\widehat{M} = \breve{A}\breve{B}^T$.  Letting $\boldsymbol{\mu_{A}}, \boldsymbol{\mu_{B}} \in \mathbb{R}^k$ contain the columns means of $\breve{A}$ and $\breve{B}$ respectively, $\widehat{M}$ can be expressed:
\begin{equation}\widehat{M} = \mu_{AB}\boldsymbol{1_{n}}\boldsymbol{1_{n}}^T +  \boldsymbol{\tilde{a}}\boldsymbol{1_{n}}^T + \boldsymbol{1_{n}}\boldsymbol{\tilde{b}}^T + \widetilde{A}\widetilde{B}^{T}, \label{svd}
\end{equation}
\vspace{-.5in}
\begin{align*}
\widetilde{A} &= (\breve{A} -  \boldsymbol{1_{n}}\boldsymbol{\mu_{A}}^T), \hspace{.5in} \boldsymbol{\tilde{a}} = \widetilde{A}\boldsymbol{\mu_{B}}, \hspace{.5in} \mu_{AB} =\boldsymbol{\mu_{A}}^T\boldsymbol{\mu_{B}},\\
\widetilde{B} &= (\breve{B} -  \boldsymbol{1_{n}}\boldsymbol{\mu_{B}}^T), \hspace{.5in} \boldsymbol{\tilde{b}} = \widetilde{B}\boldsymbol{\mu_{A}},
\end{align*}
where $\widetilde{A}$ and $\widetilde{B}$ represent multiplicative factors with mean-zero columns, $\boldsymbol{\tilde{a}}$ and $\boldsymbol{\tilde{b}}$ represent mean-zero row and column factors, and $\mu_{AB}$ is an overall mean.  

Observe that the representation in \eqref{svd} resembles that in \eqref{amenMATRIX} for the network model.  This illustrates the additive and multiplicative effects structure in the network model is similar to a rank-$k$ matrix approximation of the network.  
Note that in the decomposition in \eqref{svd}, there is functional dependence between the overall mean, additive, and multiplicative effects.  Since there are no such restrictions in the network model, the latent network effects represent a slightly larger class of approximations than the set of matrices with rank-$k$.  

From the representation in \eqref{amenMATRIX}, it is evident that the multiplicative network effects individually are nonidentifiable: the probability model for $Y$ is the same with multiplicative latent factors $U$ and $V$ as it is with factors $UG^{T}$ and $VG^{-1}$ for any nonsingular $(k \times k)$ matrix $G$.  This issue is discussed further in Sections \ref{sec:test} and \ref{sec:joint}. \\

\noindent \textbf{Non-continuous network measures}\\
Observed network information is often not continuous.  For example it is common for network data to be binary where $y_{i,j}$ is an indicator of whether the relation between nodes $i$ and $j$ exceeds some threshold, or ordinal where $y_{i,j}$ represents, for instance, the relative rank of node $j$ from the perspective of node $i$.  To model non-continuous relations, the network model in \eqref{amen} can be incorporated into a generalized linear model framework by modeling $y_{i,j} = \ell(z_{i,j})$ where $z_{i,j}$ is a continuous measure of the pairwise relation and $\ell$ is a link function defining the relationship between $z_{i,j}$ and $y_{i,j}$.  The latent continuous network measure $z_{i,j}$ is then modeled using the network model in \eqref{amen} in place of $y_{i,j}$.  In the case of binary data, a probit or logit link function may be appropriate, and in the ordinal case the ordered probit can be considered.  \cite{Hoff2012} discusses additional link functions which account for censoring of binary and ordinal relations when nodes are restricted on the number of relations they can send (i.e. the number of non-zero relations in a row of $Y$).  Section \ref{sec:app} illustrates the use of an appropriate link function for fixed rank nomination data from the National Longitudinal Study of Adolescent Health.

\section{Testing for dependencies} \label{sec:test}
The goals in an analysis of network and attribute data are often threefold: 1)  to determine whether dependencies exist between the network and attributes, 2) to model and estimate these dependencies, and finally 3) to make inference and possibly make predictions for missing data.  The first step in any such analysis is to formally test for dependencies between the network and attributes.  
   
A classical approach to determining whether there is an association between the nodal attributes $X_{n \times p}$ and network relations $Y_{n \times n}$ would be to test whether dependencies exist between $X$ and the rows of $Y$ or between $X$ and the columns of $Y$.  This would involve hypothesizing that each attribute is uncorrelated with each node's outgoing relations (H$_{0}$: Cov$(X[,i],Y[j,])=0$ for all $i$,$j$) or incoming relations (H$_{0}$: Cov$(X[,i],Y[,j])=0$ for all $i$,$j$) and investigating the evidence against these claims.  However, conventional multivariate analysis tests are not applicable to these problems since these tests address relationships between $p+n$ variables based on $n$ observations.  

We propose an alternative testing approach using the the estimated latent network factors $N_{n \times (2k+2)}=[\boldsymbol{a},\boldsymbol{b},U,V]$ from the network model in \eqref{amen}.  The nodal attributes $X=[\boldsymbol{x}_{1},...,\boldsymbol{x}_{n}]^{T}$ are independent of the network $Y$ if and only if the attributes are independent of any function of the network.  As described in Section \ref{sec:redrep}, the network factors $N$ provide a simplified representation of the network.  
Thus, we propose testing for dependence between the latent network factors $N$ and attributes $X$ on the basis that rejecting such a test would imply dependence between the network $Y$ and attributes $X$ (see Figure \ref{conceptpic}).  However, the latent network factors $N$ are not observed so in practice they must be estimated from the observed network $Y$.  In this section we propose a test for dependence between the estimated network factors and attributes, discuss invariances in the test, and describe an exact likelihood ratio testing procedure.  We also discuss alternative interpretations of the test that do not involve distributional assumptions on both the latent factors and attributes.

Suppose the nodal attributes $\{\boldsymbol{x}_{i} : i \in \{1,...,n\}\}$ are continuous and mean-zero, and let $\boldsymbol{n}_{i} = (a_{i},b_{i}, \boldsymbol{u}_{i}^{T}, \boldsymbol{v}_{i}^{T})^{T}$ denote the (estimated) latent network factors for node $i$.  We propose testing for linear dependence between the network factors and attributes using a classical multivariate test based on the assumption that the network factors and attributes are samples from a multivariate normal distribution: 
\begin{equation} \hspace{-.05in} \left( \boldsymbol{x}_{i}^{T},a_{i},b_{i}, \boldsymbol{u}_{i}^{T}, \boldsymbol{v}_{i}^{T}  \right)^{T} = \left(\boldsymbol{x}_{i}^{T}, \boldsymbol{n}_{i}^{T}  \right)^{T} \iid \text{normal}_{p+2+2k} \left(\left( \begin{array}{c}
\boldsymbol{0}\\
\boldsymbol{0} \\
 \end{array} \right),\Sigma_{XN} = \left( \begin{array}{cc}
\Sigma_{X} & \Sigma_{X,N}\\
 \Sigma_{N,X} & \Sigma_{N} \\
 \end{array} \right) \right). \label{jointxn}
 \end{equation}
The null and alternative hypotheses for this test are
\begin{equation}
\text{H}_{0}:  \Sigma_{X,N}=0 \hspace{.25in} \text{ vs. } \hspace{.25in} \text{H}_{1}:  \Sigma_{X,N}\not=0  \hspace{.25in} \text{based on \eqref{jointxn}}. \label{test}
\end{equation} 

\noindent\textbf{Network model and test invariances}\\
As mentioned in Section \ref{sec:redrep}, the network model in \eqref{amen} is invariant under transformations of the multiplicative latent factors.  Formally, 
this nonidentifiability can be expressed as a invariance of the probability model under transformations of network factors $\{\boldsymbol{n}_{i}: i \in \{1,...,n\} \}$ by elements of group  \[ \mathcal{G}_{N} = \left\{ \boldsymbol{G}_{N} =\left( \begin{array}{ccc}
 \text{I}_{2} & 0  & 0\\
0 & A^{T} & 0\\
0&0 & A^{-1} \\
 \end{array} \right): A_{k \times k} \text{ nonsingular} \right\},\]
which act via multiplication on the left: $ \boldsymbol{n}_{i} = (a_{i}, b_{i}, \boldsymbol{u}_{i}^{T},\boldsymbol{v}_{i}^{T})^{T} \rightarrow \boldsymbol{G}_{N}\boldsymbol{n}_{i} = (a_{i}, b_{i}, A^{T}\boldsymbol{u}^{T}_{i},A^{-1}\boldsymbol{v}^{T}_{i})^{T}$.  It would be undesirable for the test in \eqref{test} to depend on which 
 latent factors in the set $\{\{\boldsymbol{G}_{N}\boldsymbol{n}_{i} : i \in \{1,...,n\}\} : \boldsymbol{G}_{N} \in \mathcal{G}_{N}\}$ are selected to represent the network.  Define $\mathcal{G}$ to be the extension of group $\mathcal{G}_{N}$ to transformations of $\left(\boldsymbol{x}_{i}^{T}, \boldsymbol{n}_{i}^{T}  \right)^{T}$: \[ \mathcal{G} = \left\{ \boldsymbol{G}=\left( \begin{array}{cc}
I_{p} & 0\\
0 & \boldsymbol{G}_{N} \\
 \end{array} \right): \boldsymbol{G}_{N} \in \mathcal{G}_{N} \right\},\]  
which acts via left multiplication and leaves $\boldsymbol{x}_{i}$ unchanged.  We define $\mathcal{G}$ in order to relate the invariance in the network model parameterization to the test in \eqref{test}.

 The testing problem in \eqref{test} is itself invariant under left multiplication of $\left(\boldsymbol{x}_{i}^{T}, \boldsymbol{n}_{i}^{T}  \right)^{T}$ by elements in the group $\mathcal{F}$, where $\mathcal{F}$ is defined  \[ \mathcal{F} = \left\{ \boldsymbol{F} =\left( \begin{array}{cc}
B^{X} & 0  \\
0 & B^{N} \\
 \end{array} \right): B^{X}_{p \times p}, B^{N}_{(2k + 2) \times (2k+2)} \text{ nonsingular} \right\}.\]  
An $\mathcal{F}$-invariant test is a test for \eqref{test} that produces the same results for all attributes and network factors that are equivalent under group $\mathcal{F}$.  Observe that $\mathcal{G}$ is a subgroup of $\mathcal{F}$. This implies that an $\mathcal{F}$-invariant test will also respect the $\mathcal{G}$-invariances in the network relations probability model.  In other words, all attributes and latent network factors that are equivalent under group $\mathcal{G}$ will generate the same test results for \eqref{test} under an $\mathcal{F}$-invariant test.\\
 
 \noindent\textbf{Likelihood ratio test}\\
There is no uniformly most powerful invariant test for \eqref{test}, however the likelihood ratio test is $\mathcal{F}$-invariant, unbiased (\cite{PerlmanOlkin1980}), and generally performs well.  Let $N=[\boldsymbol{a},\boldsymbol{b},U,V]$ be the $(n \times (2k+2))$ matrix of network factors.  The likelihood ratio test statistic can be written 
\begin{equation}
\Lambda = \frac{\max_{\Sigma} \; L(\Sigma|N,X)}{\max_{\Sigma_{X},\Sigma_{N}} \; L_{0}(\Sigma_{X},\Sigma_{N}|N,X)} =  \prod_{i=1}^{p \wedge (2k+2)} (1-r_{i}^{2})^{-n/2} \label{lrt}
\end{equation}
where $L_{0}$ and $L$ refer to the likelihood corresponding to the multivariate normal model in \eqref{jointxn} with and without restricting $\Sigma_{N,X}=0$.  The term $r_{i}^{2}$ is the $i$th eigenvalue of $$(X^{T}X)^{-1/2}(X^{T}N)(N^{T}N)^{-1}(N^{T}X)(X^{T}X)^{-1/2},$$ and its positive square root is commonly referred to as the $i$th canonical correlation between $N$ and $X$.  This correlation represents the largest correlation obtainable between a linear combination of attributes and a linear combination of the network factors such that the linear combinations are uncorrelated with the respective combinations used to obtain the first $i-1$ correlations. 
 The test based on \eqref{lrt} rejects the null hypothesis for large values of $\Lambda$ and was shown to have monotonically increasing power as a function of each population canonical correlation (\cite{AndersonDasGupta1964}).

Under the null hypothesis, $W = \Lambda^{-2/n}$ has a Wilks' Lambda $U(p,2k+2,n-(2k+2))$ distribution, which is equivalent to the product of independent, Beta distributed random variables (\cite{Muirhead1982}):
$$ W \sim U(p,2k+2,n-(2k+2))= \prod_{i=1}^{p} \text{Beta} \left(\frac{n-(2k+2)-p+i}{2},\frac{2k+2}{2} \right).$$
The $\alpha$-quantiles for this distribution can be obtained via Monte Carlo estimation and used to perform exact level-$\alpha$ tests for \eqref{test}. \\ 

\noindent\textbf{Alternative interpretation of the test}\\
The test in \eqref{test} was derived as the likelihood ratio test for a model where both the network factors and attributes are samples from a normal distribution.  However in some cases these assumptions may not be appropriate.  Fortunately, alternative interpretations of the test exist that do not rely on such assumptions.  The likelihood ratio test in \eqref{lrt} for the test in \eqref{test} is the same as the likelihood ratio test to determine whether the coefficients in a linear regression are nonzero, where either the network factors are regressed on the attributes or the attributes are regressed on the network factors.  These conditional tests can be expressed 
\begin{align}
&\text{H}_{0}:  \boldsymbol{\beta}_{X|N}=\boldsymbol{0} \hspace{.05in} \text{ vs. } \hspace{.1in} \text{H}_{1}:  \boldsymbol{\beta}_{X|N}\not=\boldsymbol{0} \hspace{.2in} \text{ based on } \hspace{.2in} 
 \boldsymbol{x}_{i} | \boldsymbol{n}_{i}\iid \text{normal} \left( \beta_{X|N} \boldsymbol{n}_{i},\Sigma_{X|N} \right) \label{testcondxgn}, \text{ and }\\
&\text{H}_{0}:  \boldsymbol{\beta}_{N|X}=\boldsymbol{0} \hspace{.05in} \text{ vs. } \hspace{.1in} \text{H}_{1}:  \boldsymbol{\beta}_{N|X}\not=\boldsymbol{0}  \hspace{.2in} \text{ based on } \hspace{.2in}
 \boldsymbol{n}_{i} | \boldsymbol{x}_{i}  \iid \text{normal} \left( \beta_{N|X}\boldsymbol{x}_{i}, \Sigma_{N|X}  \right), \label{testcondngx}
 \end{align} 
where $\beta_{X|N}$ 
 and $\beta_{N|X}$ 
  are $(p \times (2+2k))$ and $((2+2k) \times p)$ matrices, respectively.  If the nodal attributes were specified as part of the study design or are binary or ordinal, it may be inappropriate to model them as Gaussian as is done in \eqref{test}.  Instead it may be preferable to test for dependence between the attributes and network factors via the conditional formulation in \eqref{testcondngx}, where no distributional assumptions are placed on $X$.  The likelihood ratio test for the tests in  \eqref{test}, \eqref{testcondxgn} and \eqref{testcondngx} are identical, so the testing framework presented here is appropriate if the assumption of normality is reasonable for one or both of the network factors and attributes.

\section{Simulation study}\label{sec:sim}
To analyze data with the test outlined in Section \ref{sec:test}, the network latent factors $N$ must be estimated from the observed network $Y$.  We expect this to result in a decrease in the power of the test in \eqref{test} compared to if the network factors were able to be observed directly.
Furthermore, we expect a greater decrease in power when the observed network relations are less informative (i.e. binary rather than continuous).  In this section we present a simulation study that quantifies the degree to which power is lost when the network factors are not observed and must be estimated from observed network relations.

Consider the network model in \eqref{amen} with one multiplicative effect ($k=1$), zero mean ($\mu=0$), and independent standard normal errors ($\rho=0$, $\sigma^{2}_{e} = 1$):
\begin{align}
y_{i,j} &= a_i + b_j + u_{i}v_{j} + e_{i,j},  \hspace{.5in} a_{i},b_{j},u_{i},v_{j} \in \mathbb{R},  \hspace{.5in}  e_{i,j} \sim \text{normal}(0,1).
\end{align}
We consider the case where one nodal attribute is of interest ($p=1$) and  the attribute and latent network factors have one of the following covariance structures:  
\begin{enumerate}[\indent A)]
\item
 Cov[$(x_{i},a_{i}, b_{i}, {u}_{i},{v}_{i})$] $= \Sigma_{XN} = $ I $ + \; \gamma E_{x,a}$,
 \item
 Cov[$(x_{i},a_{i}, b_{i}, {u}_{i},{v}_{i})$] $= \Sigma_{XN} = $ I $ + \; \gamma E_{x,u}$.
 \end{enumerate}
$E_{x,a}$ is the ($ 5 \times 5$) matrix of zeros with a one in the entires corresponding to Cov[$x,a$]  and Cov[$a,x$], and $E_{x,u}$ is defined analogously.  In scenario A the attribute and each network factor are uncorrelated, except the additive sender factor $a_{i}$ and the attribute $x_{i}$ which have correlation $\gamma$.  Similarly, in scenario B correlation $\gamma$ exists between the sender multiplicative factor $u_{i}$ and the attribute $x_{i}$.    

Monte Carlo estimates of the power based on the level-$0.05$ likelihood ratio test in \eqref{lrt} for the test in \eqref{test} were computed for squared correlation values $\gamma^{2} \in \{-0.05, 0, 0.05, 0.1, 0.15, 0.2\}$, network sizes $n \in \{25, 50, 100\}$ and three decreasingly informative observations of the network:\\
\indent 1. $N =[ \boldsymbol{a}, \boldsymbol{b}, U, V ]$ is observed;\\
\indent 2. $N$ is estimated from a continuous network $Y$ according to \eqref{amen};\\
\indent 3. $N$ is estimated from a binary network $B_{d}$, where $B_{d}$ is defined as $B_{d} = \{b_{i,j}: b_{i,j}=1 \text{ if } y_{i,j} > y_{d},\; 0 \text{ otherwise}\}$ and $y_{d}$ is chosen such that the proportion of network relations greater than $y_{d}$ (i.e. the network density) is $d$. \\
Notice that the binary network $B_{d}$ is a deterministic function of the continuous network $Y$.  We consider the binary networks with density $0.5$ and $0.15$.  The former case represents a relatively dense binary network with many observed relations, whereas the latter case reflects more common network seen in survey data where information about only a small number of nodes' relations are available.  For the continuous network $Y$ and binary networks $B_{d}$, the Bayesian estimation procedure was used to obtain estimates of the latent network factors.  A probit link function was specified for the binary networks.  The additive factors $\boldsymbol{a}$ and $\boldsymbol{b}$ were estimated by their posterior means, and the multiplicative factors $U$ and $V$ were estimated by the first left and right singular vector of the posterior mean of the multiplicative effect $UV^{T}$. 

\begin{figure}
\begin{center}
\includegraphics[scale=.86]{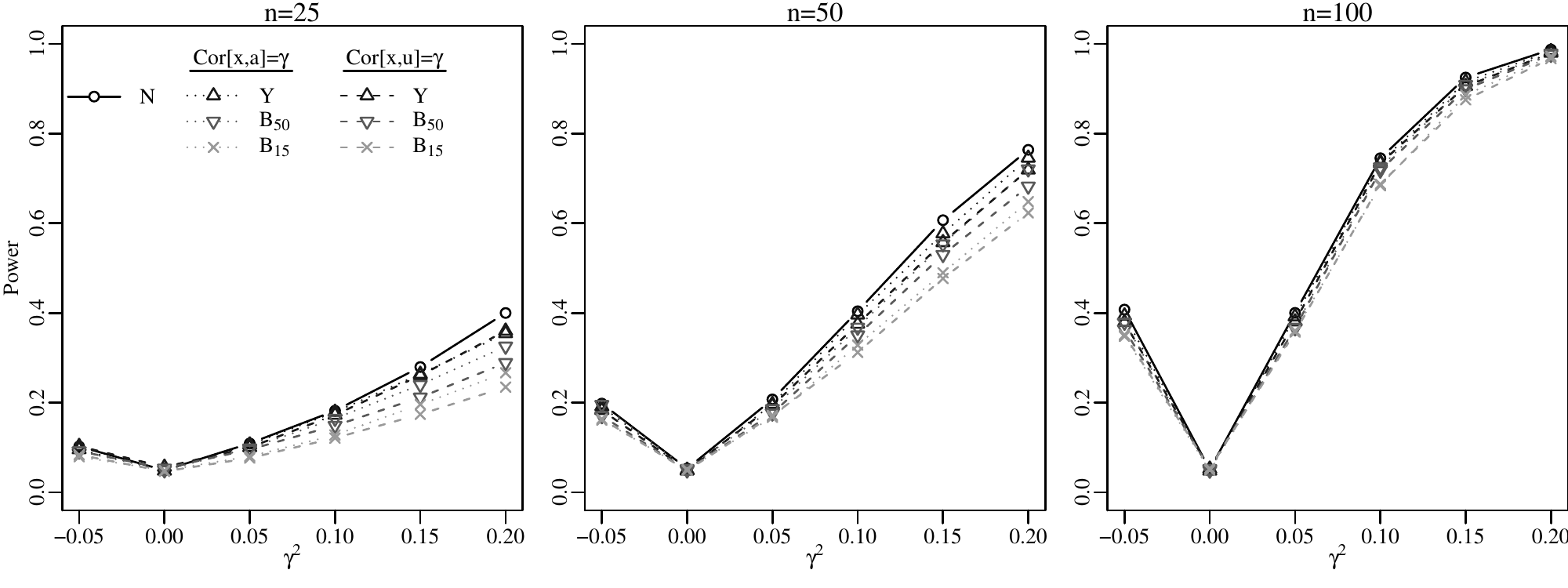}
\caption{Power when testing for independence between a single attribute $x_{i}$ and network factors $\{a_{i},b_{i},u_{i},v_{i}\}$ based on four types of network observations (latent network factors $N$, continuous network $Y$,  binary network $B_{0.50}$, binary network $B_{0.15}$).}
\label{sim}
\end{center}
\end{figure}

Figure \ref{sim} shows the power estimates for the two covariance structures A and B and the four network observations ($N$, $Y$, $B_{0.5}$, $B_{0.15}$).  A single power curve is shown for the latent network factors $N$ since the correlation structures A and B are equivalent with respect to the invariances of the test in \eqref{test}. Most notably, Figure \ref{sim} illustrates there is relatively little power lost when the network factors are estimated from an observed continuous or binary network, even when the network size is small.  The power of the test is slightly larger when dependence exists between the attribute and an additive factor compared to when it exists between the attribute and a multiplicative factor for continuous and binary network observations.  This is likely a consequence of the relative ease with which additive effects are estimated compared to interaction effects.  As expected, the power of the test decreases as the observed network information becomes less informative ($N  \rightarrow Y \rightarrow B_{0.5} \rightarrow B_{0.15}$), although for even moderate network sizes the power loss is negligible.

\section{Joint model for the network and nodal attributes}\label{sec:joint}
If the test in Section \ref{sec:test} rejects the null hypothesis of independence between the attributes and network factors, there is often interest in estimating and making inference on the dependencies, as well as predicting missing network and attribute information.  Addressing such inference objectives requires joint modeling of the network $Y$ and attributes $X$. We propose jointly modeling the network $Y$ and attributes $X$ via a model composed of the network relations model in \eqref{amen} and \eqref{err}, and the latent factor and attribute model in \eqref{jointxn}.   For completeness, we include all components of the joint model below. 
\vspace{-.2in}
\begin{align}
 y_{i,j} &= \mu + a_i + b_j + \boldsymbol{u}_{i}^{T}\boldsymbol{v}_{j} + e_{i,j}\label{jointmodel_net}\\
(e_{i,j},e_{j,i})^T& \iid \text{ normal}_{2}\Big(\boldsymbol{0},\sigma^{2}_{e}\Big(\begin{smallmatrix}
1&\rho\\ \rho&1
\end{smallmatrix} \Big) \Big) \label{jointmodel} \\
\left( \boldsymbol{x}_{i}^{T},a_{i},b_{i}, \boldsymbol{u}_{i}^{T}, \boldsymbol{v}_{i}^{T}  \right)^{T} &= \left(\boldsymbol{x}_{i}^{T}, \boldsymbol{n}_{i}^{T}  \right)^{T} \iid \text{normal}_{p+2+2k} \Big(\Big( \begin{array}{c}
\boldsymbol{0}\\[-8pt]
\boldsymbol{0} 
 \end{array} \Big), \;\Sigma_{XN} = \Big( \begin{array}{cc}
\Sigma_{X} & \Sigma_{X,N}\\[-8pt]
 \Sigma_{N,X} & \Sigma_{N} 
 \end{array} \Big) \Big) \label{jointmodel_lat}
\end{align}\\[-23pt]
Inference for the dependence and conditional dependencies between the attributes and network is based on the covariance matrix $\Sigma_{XN}$.\\

\noindent \textbf{Simplified parameterization}\\
The nonidentifiability of the latent factors discussed in Sections \ref{sec:redrep} and \ref{sec:test} translates to nonidentifiability of portions of the covariance matrix $\Sigma_{XN}$.  However, by restricting the covariance matrix to have specific structure, the $\mathcal{G}$-invariance of the network model due to the multiplicative latent factors can be removed.

We propose reparameterizing the model for the latent factors and attributes in \eqref{jointmodel_lat} by
\begin{equation}
 \left( \boldsymbol{x}_{i}^{T},a_{i},b_{i}, \boldsymbol{u}_{i}^{T}, \boldsymbol{v}_{i}^{T}  \right)  \iid \text{normal}_{p+2+2k} \left(\left( \begin{array}{c}
\boldsymbol{0}\\
\boldsymbol{0} \\
\boldsymbol{0} \\
 \end{array} \right),\Sigma_{XN} = \left( \begin{array}{ccc}
 \Sigma_{Xab}& \Sigma_{Xab,U}& \Sigma_{Xab,V}\\
 \Sigma_{U,Xab}& D& \Sigma_{U,V}\\
 \Sigma_{V,Xab}& \Sigma_{V,U}& D\\
 \end{array} \right) \right), \label{sparam}
 \end{equation}
 where $D$ is a diagonal matrix with decreasing elements along the diagonal.  This joint model defined by \eqref{jointmodel_net}, \eqref{jointmodel}, and \eqref{sparam} is not invariant to transformations of the network factors and attributes by elements in the group $\mathcal{G}$, however it continues to posses non-identifiabilty with respect to signs of the entries in $U$ and $V$.  Specifically, the probability of the observed network Y and attributes X is the same with parameters $\{U,V,\Sigma_{XN}\}$ as it is with parameters $$\{US,VS,\Big(\begin{smallmatrix}
I_{p+2}&0 &0 \\ 0&S & 0 \\ 0 & 0 & S
\end{smallmatrix} \Big)\Sigma_{XN}\Big(\begin{smallmatrix}
I_{p+2}&0 &0 \\ 0&S & 0 \\ 0 & 0 & S
\end{smallmatrix} \Big)\},$$ where $S_{k \times k}$ is any diagonal matrix with $\pm1$'s along the diagonal.\\

 \noindent \textbf{Relation to reduced rank regression}\\ 
The expectation of the network relations conditional on the attributes based on \eqref{jointmodel_net} resembles that of a reduced rank regression on (multiplicative) attribute interaction effects.  This is noteworthy as the motivations underlying reduced rank regression parallel many of the arguments supporting this network modeling framework.

The expectation of the network factors conditional on the attributes can be written $$\text{E}[(a_{i},b_{i},\boldsymbol{u}_{i}^{T}, \boldsymbol{v}_{i}^{T})^{T}]=(\beta_{a|X}\boldsymbol{x}_{i}, \beta_{b|X}\boldsymbol{x}_{i},(\beta_{U|X}\boldsymbol{x}_{i})^{T},(\beta_{V|X}\boldsymbol{x}_{i})^{T})^{T},$$ 
where $\beta_{a|X}$, $\beta_{b|X}$ are $(p \times 1)$ vectors and $\beta_{U|X}$ and $\beta_{V|X}$ are $((2+2k) \times p)$ matrices of coefficients based on $\Sigma_{XN}$.  Since the latent factors for different nodes are modeled as independent, the expectation of the network relations in \eqref{jointmodel_net} conditional on the attributes is 
$$ \text{E}[y_{i,j}|\boldsymbol{x}_{i},\boldsymbol{x}_{j}]  = \mu + \beta_{a|X}\boldsymbol{x}_{i} + \beta_{b|X}\boldsymbol{x}_{j} + \boldsymbol{x}_{i}^{T}\beta_{U|X}^{T}\beta_{V|X}\boldsymbol{x}_{j}.$$
The interaction term $\boldsymbol{x}_{i}^{T}\beta_{U|X}^{T}\beta_{V|X}\boldsymbol{x}_{j} $ represents a linear combination of all possible pairwise products between the $p$ sender and $p$ receiver attributes, resulting in $p^{2}$ linear effects.  The coefficients on these linear effects are given by the $(k \times k)$ matrix $\beta_{U|X}^{T}\beta_{V|X}$, whose rank is at most equal to the minimum of $p$ and $k$.  Therefore, if the number of attributes is greater than the number of multiplicative network factors ($p \ge k$), linear constraints will exist among the $p^{2}$ effect coefficients.  In reduced rank regression the coefficient matrix corresponding to the regression of a multivariate outcome on a multivariate predictor is restricted to be reduced rank (\cite{Anderson}, see \cite{ReinselVelu} for a comprehensive review).  
This approach to parameter dimension reduction is motivated by improvement in parameter estimation and interpretation.  A similar goal exists in network modeling and is achieved here using the latent network factors.  Modeling dependencies between the latent network factors $N$ and attributes $X$ instead of between the network relations $Y$ and attributes $X$ directly allows us to parsimoniously estimate and characterize complex (multiplicative) dependencies without defining a complicated regression model for the network relations.  This approach is especially advantageous when the number of attributes is large and/or it is likely at most a small number of attribute pairs are related to the network.    \\   
 
 \noindent \textbf{Estimation}\\ 
Estimation of the parameters in the joint network and attribute model is straightforward in a Bayesian context, where inference is based on the joint posterior distribution of the network factors $\{ \boldsymbol{a}, \boldsymbol{b}, U, V\}$ and parameters $\{\sigma^{2}_{e}, \rho, \Sigma_{XN}\}$ given the data $\{X, Y\}$.  Since an analytic expression of the posterior distribution is not available, it is approximated by samples generated from a Markov chain Monte Carlo (MCMC) algorithm.  The MCMC procedure implemented in the \texttt{R} package `amen' for the model described in Section \ref{sec:redrep}, where the additive and multiplicative factors are uncorrelated, was adapted for the joint model presented here.  Details regarding the families of prior distributions considered and the corresponding MCMC algorithm are included in the appendix.  Code is provided at the corresponding author's website.

\section{Analysis of AddHealth data}\label{sec:app}

We consider data from a survey of 389 high-school students from the National Longitudinal Study of Adolescent Health (AddHealth) (\cite{AddHealth}) and investigate whether evidence exists that student friendships are related to student health behaviors and grade point average (GPA).  The data we use includes same-sex friendship nomination data, whereby students identified the top five friends of their sex, in addition to demographic and behavioral information.  The data considered here can be described as follows:
\begin{itemize}
\item
\textbf{network information -} $R = \{ r_{i,j} \}$: $r_{i,j}$ is the rank of student $j$ in student $i$'s listing of friends (5 = highest, 1 = lowest) or $0$ if student $i$ did not list student $j$;
\item
\textbf{nodal attributes - }$X = [\boldsymbol{x}^{\text{exercise}},\boldsymbol{x}^{\text{drink}},\boldsymbol{x}^{\text{gpa}}]$: standardized measures of exercise frequency, drinking frequency, and grade point average;
\item
\textbf{nodal covariate - }$W = [\boldsymbol{w}^{\text{grade}}]$: student grade level (9, 10, 11, or 12).
\end{itemize}

Students in the same grade and adjacent grades are more likely to be friends than students many grades apart.  For this reason, we refine our question of interest to be whether students' attributes (exercise, drinking, and GPA) are associated with their network relations' while controlling for their grade.    
  
We use the fixed rank nomination likelihood introduced in \cite{Hoff2012} to model the observed network ranks and restriction that at most five friends could be listed on the survey.  This likelihood assumes each observed network relation $r_{i,j}$ is the function of an underlying (latent) continuous measure $z_{i,j}$ such that the following relation consistencies are satisfied:
\begin{align}
&r_{i,j} > 0 \Rightarrow z_{i,j} > 0,\notag \\
&r_{i,j} > r_{i,k} \Rightarrow z_{i,j} > z_{i,k},\label{FRN}\\
&r_{i,j} = 0 \text{ and student $i$ listed $<5$ friends} \Rightarrow z_{i,j} \le 0. \notag
\end{align}
The first association is the link function used in probit regression which assumes that if a friendship is reported, the latent friendship value must exceed a given threshold (in this case 0).  The second relation assures consistency of the ranks with the latent friendship measures.  Finally, the last association posits that friendships between a given student and all students he/she did not list as a friend must be below the friendship threshold if the nominating student listed fewer than five friends. 

The network model for the latent relations $z_{i,j}$ is that given in \eqref{amen} with additional regression terms for whether students are in the same grade $w^{s}_{i,j}$ and whether they are in adjacent grades $w^{a}_{i,j} $:   
\begin{align}
z_{i,j} = \mu + \beta_{s}w^{s}_{i,j} + \beta_{a}w^{a}_{i,j}  + a_i + b_j + \boldsymbol{u}_{i}^{T}\boldsymbol{v}_{j} + e_{i,j},  \hspace{.5in} a_{i},b_{i} \in \mathbb{R}, \hspace{.5in} \boldsymbol{u}_{i},\boldsymbol{v}_{i} \in \mathbb{R}^{k}.
 \label{Addnet}
\end{align}\\[-25pt]

\begin{figure}
\begin{center}
\includegraphics[scale=.9]{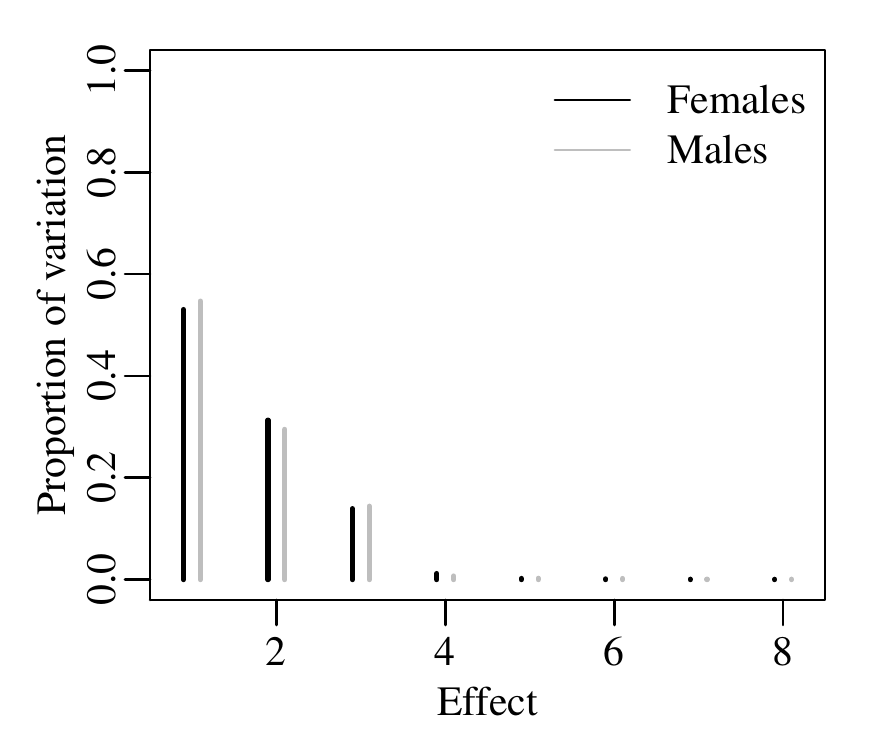}
\caption{Proportion of variation in the posterior mean eight factor multiplicative effect $\widehat{M}$ that is explained by each multiplicative effect.}
\label{AddRank1}
\end{center}
\end{figure}

\noindent \textbf{Selection of factor dimension $k$}\\
The multiplicative factor dimensions $k$ for the male and female networks were determined using a method analogous to the scree plot method which is commonly used in factor analysis and principal components analysis.  The network model in \eqref{FRN} and \eqref{Addnet} was fit to each gender network with $k=8$.  Let $M$ denote the posterior mean estimate of the multiplicative network effect $UV^T$, and $\widehat{M}$ represent the rank eight matrix approximation of $M$ based on the singular value decomposition.  The total variation in $\widehat{M}$ is equal to the sum of the squared singular values: $ ||\widehat{M}||_{F}^{2} = \sum_{\ell = 1}^{8} \lambda_\ell^2$, where $||\cdot||_{F}$ denotes the Frobenius norm and $\lambda_{i}$ is the $i$th singular value.  Figure \ref{AddRank1} shows the proportion of the total variation in $\widehat{M}$ attributed to each multiplicative effect (i.e. $\lambda_{i}^2 / \sum_{\ell = 1}^{8} \lambda_\ell^2$).  For both the male and female network the large majority of the variation in the network relations explained by the eight multiplicative effects is associated with the first three effects.  Thus, the multiplicative effect dimension was selected to be three for both networks.\\  

\noindent \textbf{Testing for dependence}\\
As discussed in the Introduction, a traditional approach to modeling dependence between the network relations and nodal attributes would be to include regression terms in the form of sender, receiver, and interaction effects for the three attributes.  Including all such effects using this approach would require 15 regression terms.  However, by performing the test of independence proposed in Section \ref{sec:test} based on the latent network factors, we are able to assess the evidence for any relationship between the attributes and network without creating a potentially unnecessarily complex network model or performing any model selection.
 
The latent network factors for the model in \eqref{FRN} and \eqref{Addnet} with $k=3$ were estimated for the male and female networks.  The additive factors $\boldsymbol{a}$ and $\boldsymbol{b}$ were estimated by their posterior means, and the multiplicative factors $U$ and $V$ were estimated by the first three left and right singular vectors of the posterior mean of the multiplicative effect $UV^{T}$.  The test of independence between the network factors and the three nodal attributes for the female and male network resulted in $p$-values $<0.001$.  Therefore, based on a $0.05$ level test, we reject the null hypothesis of independence between the student attributes and their network relations after accounting for grade structure. \\

\noindent \textbf{Jointly modeling the network and attributes}\\
The rejections of the tests of independence between the attributes and network suggests the network factors are informative for nodal attribute data.  To investigate this claim we performed a 20-fold cross validation on each sex dataset in which $5$\% of data for each attribute was treated as missing in each experiment.  We compared predictions for the missing attributes based on the observed attributes alone to predictions based on both the network and observed attributes.  The predictions based solely on the attributes were the fitted values from a regression of each attribute on all other attributes.  The predictions based on the network and attributes were the posterior mean estimates from the Bayesian estimation procedure for the joint network and attribute model introduced in Section \ref{sec:joint}.  For each sex dataset, a Markov chain was run for 500 iterations of burn-in followed by an additional 500,000 iterations and samples were thinned to every 25th iteration, resulting in 20,000 simulated values for each missing element.  The average effective sample size was 2,607 for the male network and 734 for the female network.  

\begin{table}[ht]
\caption{Mean squared error for predictions from 20-fold cross validation.}
\centering
\begin{tabular}{l|ccc|ccc}
& \multicolumn{3}{c|}{Males} &\multicolumn{3}{c}{Females}\\
Method & Exercise & Drinking & GPA & Exercise & Drinking & GPA \\ 
\hline \hline
Regression (attributes only) & 1.89 & 3.24 &  2.38 & 1.67& 2.38 & 2.29 \\
Joint model (attributes \& network) &1.75 & 2.69 & 2.18  &1.61 & 2.17 &1.93 \\ 
\hline
\% improvement  & 7.4 & 17.0 & 8.4 & 3.6 & 8.8 & 15.7 \\
\end{tabular}
\label{cv}
\end{table}

Table \ref{cv} shows the mean squared error over the 20 cross validations for each attribute and each sex dataset.  The predictions based on the network and attributes improved upon the predictions based on the attributes alone for both sexes and all attributes.  The improvement was greatest for male drinking frequency and female GPA where prediction mean squared error was reduced by about 15\%.  This illustrates that when dependence exists between the network and attributes, improvements in the predictions of missing values can be obtained by using both the network and attribute information.

\section{Discussion}\label{sec:dis}

In this article we introduced an approach for testing whether dependencies exist between a network and attribute data that relies on a simplified representation of the network in terms of latent node-specific factors.  The proposed method tests for dependencies between the network latent factors and attributes as a surrogate for testing for dependencies between the network and attributes.  This test was shown to have the correct level under the null hypothesis of independence and have only a slight loss in power due to the fact that the network factors are not directly observed.  Methodology for jointly modeling the network and attributes was also introduced, and in a cross validation experiment, we illustrated that predictions for missing attributes can be improved by basing predictions on both observed network and attribute information rather than on attribute information alone.  

A historically difficult problem not addressed here is how to select the number of multiplicative factors for the network model.  In Section \ref{sec:app} we illustrated a procedure similar to the scree plot method used frequently to choose the number of factors in factor analysis and the number of eigenvectors in principal component analysis.  An alternative approach would be to incorporate the dimension selection into the model by placing a prior on the number of factors similar to that proposed in \cite{Hoff2007} for the singular value decomposition.  However this would greatly increase the complexity of the model and computation time of estimation.

\bibliographystyle{apa} 
\bibliography{net} 

\appendix
\section{Bayesian estimation procedure}
In this section we outline the Bayesian estimation procedure used to obtain parameter estimates for the joint attribute and network model in \eqref{jointmodel_lat}.  This procedure is extremely similar to that implemented in the `amen' package in the statistical computing program \texttt{R}.  We present the simple case here where the observed network $Y$ is continuous, there are no regression terms in the network model, and there is no missing data.  For details on accommodating non-continuous network data see \cite{Hoff2012} and for including regression terms see \cite{Hoff2005}.\\

\noindent Model - 
\vspace{-.2in}
\begin{align*}
y_{i,j} &= \mu + a_i + b_j + \boldsymbol{u}_{i}^{T}\boldsymbol{v}_{j} + e_{i,j},\notag\\
(e_{i,j},e_{j,i})^T& \iid \text{ normal}_{2}\left(\boldsymbol{0},\sigma^2_{e}\Big(\begin{smallmatrix}
1&\rho\\ \rho&1
\end{smallmatrix} \Big)\right) \\ 
\left( \boldsymbol{x}_{i}^{T},a_{i},b_{i}, \boldsymbol{u}_{i}^{T}, \boldsymbol{v}_{i}^{T}  \right)^{T} &
 \iid \text{normal}_{p+2+2k} \left( \boldsymbol{0}
,\Sigma_{XN} 
  \right) 
\end{align*}
\noindent \text{Prior distributions -  }
\begin{align*}
\sigma^{-2}_{e} &\sim \text{gamma}(1/2,1/2)  \\
\rho &\sim \text{uniform}(-1,1) \\
\Sigma_{XN}^{-1} & \sim \text{Wishart}\left(p+2+2k+1,\Big(\begin{smallmatrix} \Sigma_{X0}^{-1} &0\\ 0&\text{I}_{2+2k} 
\end{smallmatrix} \Big)\right) 
\end{align*}
  
\noindent Markov chain Monte Carlo algorithm -\\
\indent Given initial values of all latent variables $\{\boldsymbol{a},\boldsymbol{b},U,V\}$ and parameters $\{\Sigma_{XN}, \rho, \sigma^{2}_{e}\}$, the algorithm proceeds as follows:
 \begin{enumerate}
 \item
 Sample $\boldsymbol{a}, \boldsymbol{b} | Y, X, U, V, \Sigma_{XN}, \rho, \sigma^{2}_{e}$ (normal).
 \item
 Sample $\Sigma_{XN} | Y, X, \boldsymbol{a}, \boldsymbol{b},  U, V, \rho, \sigma^{2}_{e}$ (inverse-Wishart). 
 \item
Update $\rho$ using a Metropolis-Hastings step with proposal $\rho^{*} | \rho \sim \text{truncated normal}_{[-1,1]}(\rho,\sigma^{2}_{\rho})$; 
\item
Sample $ \sigma_{e}^{2}| Y, X, \boldsymbol{a}, \boldsymbol{b}, U, V, \rho, \Sigma_{XN} $ (inverse-gamma). 
\item
For each latent factor $i$:
\begin{itemize}
\item
Sample $U[,i] |Y, X, \boldsymbol{a}, \boldsymbol{b} , U[,-i], V, \rho, \sigma^{2}_{e} , \Sigma_{XN}$ (normal); 
\item
Sample $V[,i] | Y, X, \boldsymbol{a}, \boldsymbol{b} , U, V[,-i], \rho, \sigma^{2}_{e} , \Sigma_{XN}$ (normal).
\end{itemize}
 \end{enumerate} 
 
 Although the estimation algorithm is not constructed based on the unique parameterization of the model, each sample of network factors from the posterior distribution can be transformed using the covariance matrix $\Sigma_{XN}$ sample to represent a sample from \eqref{sparam}.  Inference for the relative likeliness of parameter values is based on the posterior distribution over the parameter equivalence classes associated with representations congruent with \eqref{sparam}.

\end{document}